\documentclass{ifacconf}

\usepackage{natbib}
\usepackage{graphicx}
\usepackage{epsfig}
\usepackage{amsmath}
\usepackage{amssymb}
\usepackage{amsfonts}

\newcommand{\q}[1]{\vert #1 \rangle}
\newcommand{\qd}[1]{\langle #1 \vert}

\newcommand{\daag}{^{\dagger}}
\newcommand{\minou}{\text{-}}
\newcommand{\ba}{\text{\bf{a}}}
\newcommand{\bn}{\text{\bf{N}}}
\newcommand{\bid}{\text{\bf{I}}}
\newcommand{\bU}{\text{\bf{U}}}

\newcommand{\targ}{\bar{n}}

\begin{document}

\begin{frontmatter}

\title{Robust open-loop stabilization of Fock states by time-varying quantum interactions\thanksref{footnoteinfo}}

\thanks[footnoteinfo]{This paper presents research results of the Belgian Network DYSCO (Dynamical Systems, Control, and Optimization), funded by the Interuniversity Attraction Poles Programme, initiated by the Belgian State, Science Policy Office. The authors were partially supported by the ANR, Projet Blanc EMAQS ANR-2011-BS01-017-01 and Projet C-QUID BLAN-3-139579.}

\author[First]{Alain Sarlette}
\author[Second]{Pierre Rouchon}

\address[First]{SYSTeMS, Ghent University, Technologiepark Zwijnaarde 914, 9052 Zwijnaarde, Belgium (e-mail: alain.sarlette@ugent.be)}
\address[Second]{Centre Automatique et Syst\`{e}mes, Mines ParisTech, 60 boulevard Saint Michel, 75006 Paris, France\\ (e-mail: pierre.rouchon@mines-paristech.fr)}


\begin{abstract}
A quantum harmonic oscillator (spring subsystem) is stabilized towards a target Fock state by reservoir engineering. This passive and open-loop stabilization works by consecutive and identical Hamiltonian interactions with auxiliary systems, here three-level atoms (the auxiliary ladder subsystem), followed by a partial trace over these auxiliary atoms. A scalar control input governs the interaction, defining which atomic transition in the ladder subsystem is in resonance with the spring subsystem. We use it to build a time-varying interaction with individual atoms, that combines three non-commuting steps. We show that the resulting reservoir robustly stabilizes any initial spring state distributed between $0$ and  $4\targ+3$ quanta of vibrations towards a pure target Fock state of vibration number $\targ$. The convergence proof relies on the construction of a strict Lyapunov function for the Kraus map induced by this reservoir setting on the spring subsystem. Simulations with realistic parameters corresponding to the quantum electrodynamics setup at Ecole Normale Sup\'{e}rieure further illustrate the robustness of the method.
\end{abstract}

\end{frontmatter}


\section{Introduction}

The last decades have seen a surge of developments on the control of systems that feature essential quantum dynamics (see e.g.~\cite{wiseman-milburn:book}). A major motivation is to harness their peculiar possibilities for IT applications ranging from fundamentally-secure communication, over hyper-precise measurements, to the quantum computer (see e.g.~\cite{NielsenChuang,HRBook}). A basic building block for operations with quantum dynamics is the ability to produce and stabilize a wealth of different `target' states. The extreme fragility, limited measurement and control possibilities in quantum systems make this already a challenging task for many target states with non-trivial (and thus interesting) properties.

Since an isolated quantum system fundamentally follows pure Hamiltonian dynamics, asymptotic stabilization necessarily requires interaction with the external world and relies on the theory of open quantum systems. Such interactions can involve measurement and feedback action, called {\em measurement-based feedback}, or also control by tailored interaction as broadly advocated in~\cite{WillemsBeh} and called {\em coherent feedback} in the quantum context.  Measurement-based feedback introduces specific difficulties because quantum measurement is fundamentally limited to partial state knowledge and always perturbs the measured system (``back-action''). Controller design therefore always needs to follow an interactions-based reasoning (see e.g.~\cite{FeebackPhot_PRA09,FeebackPhot_Nature11}). Coherent feedback involves a specific structure, based on joint evolution on a tensor product of the interacting subsystems that leads to dissipative and/or stochastic quantum dynamics for the target subsystem (see e.g.~\cite{gough-james:ieee10,kerckhoff-et-al:PRL2010}.

\emph{Reservoir engineering} is a systematic method related to coherent feedback for \emph{open loop} stabilization of quantum systems through tailored interaction. For discrete-time quantum systems, the target system repeats the same interaction with a succession of auxiliary systems, which are discarded after interaction. The dynamics on the target system resulting from repeating the same interaction process, takes the form of a Kraus map, see \cite{Kraus83}. A random Kraus map would generally drive the system to a mixed quantum state (featuring classical uncertainties). For reservoir action, interaction with each individual auxiliary system must be tailored such that the Kraus map has a desired pure state as asymptotically stable equilibrium. General techniques have been proposed to tailor Kraus maps, provided any chosen unitary evolutions can be applied to the target subsystem by controlling its Hamiltonian (kind of `complete actuation' on the subsystem Hamiltonian, see e.g.~\cite{TicozziViola2009,BolognaniTicozzi2010}). There are however many systems where control over the Hamiltonian is instead very restricted.

This paper considers such a situation, where the target system is an electromagnetic field mode, interacting with three-level atoms as auxiliary systems. This quantum electrodynamics situation is a prototype for so-called universal spring-spin systems (see e.g.~\cite{HRBook}).  The spin-spring  Hamiltonian is controlled by shifting the frequencies of the atomic transitions as a function of time, through the Stark effect. This yields a very low-dimensional input signal to tailor the target system's evolution governed by the resulting Kraus map.

\cite{MyPRL2011} propose a method  to stabilize so-called `Schr\"odinger cat states' with this setup. It builds the overall unitary interaction operator as a symmetric product of non-commuting basic operators, by varying the input signal during the field's interaction with each atom. We prove here that stabilizing pure photon states (`Fock states') is also possible. \cite{WaltherTS} have proposed a method based on `trapping state conditions', where the auxiliary systems are two-level atoms interacting resonantly with the quantized field mode: the target Fock state is an equilibrium of the Kraus map but it is a saddle point with stable manifold in some but not all relevant directions. The present paper proposes a modification of this method that makes the target Fock state a stable equilibrium of the Kraus map in the most relevant Hilbert subspace. We therefore build an interaction that combines the `trapping state' approach of \cite{WaltherTS} with the construction of symmetric products of non-commuting operators from \cite{MyPRL2011}.

We give formal convergence properties for the resulting scheme, both in absence and in presence of a disturbing environment (Theorem~\ref{thm:conv} based on the construction of a strict Lyapunov function, and Proposition~\ref{prop:decoherence}). In section~\ref{sec:description} we define the target system, the auxiliary three-level atoms, the interaction Hamiltonian with its scalar control input $u$ and the associated Kraus map. In section~\ref{sec:control}, the operators appearing in the Kraus map are computed for our open-loop piecewise constant control~\eqref{eq:usig}. Section~\ref{sec:conv} is devoted to convergence analysis. Simulations in Section~\ref{sec:sim} briefly explore the influence of parameter uncertainties and illustrate the robustness for realistic QED parameters.


\section{System description} \label{sec:description}

We consider a quantum electrodynamics setup as described e.g.~in \cite{HRBook}. The target system is a field mode, with infinite-dimensional Hilbert space $\mathcal{H}=\text{span}(\q{0}, \ldots, \q{n},...\,)$ and free Hamiltonian
\begin{equation}\label{eq:Hf}
	H_f = \sum_{n=0}^{+\infty} n \, \omega_f \q{n}\qd{n} \;
\end{equation}
where $\omega_f > 0$ is the field mode pulsation and the index $n$ gives the photon number. This paper denotes by $\q{n}$ the pure photon number eigenstates, called \emph{Fock states}. The photon number operator is defined by $\bn = \sum_{n=0}^{+\infty}\; n \; \q{n}\qd{n} = H_f \, / \, \omega_f$. We will further use the notation
$$f_{\bn} = f(\bn) = \sum_{n=0}^{+\infty}\; f(n) \; \q{n}\qd{n} = \sum_{n=0}^{+\infty}\; f_n \; \q{n}\qd{n}$$
for an operator that is diagonal in the Fock basis, for any function\footnote{We take the convention, especially useful in quantum mechanics, that $\mathbb{N}$ includes $0$.} $f: \mathbb{N} \mapsto \mathbb{C}$. We denote by $\mathcal{H}_{n_1}^{n_2}$ the Hilbert subspace spanned by Fock states $\q{n_1},\q{n_1+1},...,\q{n_2}$. A stream of identical atoms consecutively interact with this field mode according to the Jaynes-Cummings model, playing the role of a `reservoir' to stabilize the field in a target state. We consider three atomic levels $\q{g},\q{e},\q{m}$, with free Hamiltonian
\begin{equation}\label{eq:Ha}
	H_a =  \omega_m \q{m}\qd{m} - \omega_g \q{g}\qd{g}
\end{equation}
up to an irrelevant term that is a multiple of the identity operator. The transition frequencies between levels $(\q{g},\q{e})$ and $(\q{e},\q{m})$, that is $\omega_g \in \mathbb{R}$ and $\omega_m \in \mathbb{R}$ respectively, depend on an input $u \in \mathbb{R}$. The latter represents an energy shift induced on $\q{e}$ by an external field through a Stark effect, such that
\begin{equation}\label{eq:omofu}
\omega_g(u) = \overline{\omega}_g + u \;\; \text{ and } \;\; \omega_m(u) = \overline{\omega}_m - u \; .
\end{equation}
The constants $\overline{\omega}_g$ and $\overline{\omega}_m$ are the transition pulsations in absence of external field. For simplicity, the following assumes that $\omega_g(u) > 0$ and $\omega_m(u) > 0$: $\q{g}$ corresponds to the lowest atomic level whereas $\q{e}$ and  $\q{m}$  are the first and second excited states (3-level ladder system). 	The proposed method can be adapted to other energy arrangements, like V-structures, by adding short external control pulses that switch the atomic state at specific points in our scheme.

To make atomic transitions and field mode interact, we consider the following realistic situation, in the spirit of \cite{Carvalho2011}: $\vert \overline{\omega}_g-\omega_f \vert,\vert \overline{\omega}_m -\omega_f \vert\ll \omega_f$;
$\vert u \vert\sim \vert \overline{\omega}_g-\omega_f \vert$, $\vert u \vert\sim \vert \overline{\omega}_m -\omega_f \vert$.
Then with the standard rotating wave approximation, the atom-field interaction is described by the Hamiltonian:
\begin{equation}\label{eq:Hc}
	H_c = i\, \frac{\Omega}{2} \, (\, \ba\daag\, (\q{g}\qd{e}+\q{e}\qd{m}) - \ba\, (\q{e}\qd{g}+\q{m}\qd{e}) \,) \; .
\end{equation}
Here $i = \sqrt{-1}$ and $\Omega$ is the interaction strength factor, assumed to be equal for transitions $(g,e)$ and $(e,m)$; photon annihilation operator $\ba$ is defined by
$\ba = \sum_{n=1}^{+\infty} \; \sqrt{n} \; \q{n\minou1} \qd{n}$ in the Fock basis; and $\phantom{f}\daag$ denotes the adjoint of an operator (complex conjugate transpose of the associated matrix). We have the fundamental identities $\ba\daag \ba = \bn$, as well as $\ba \, f(\bn) = f(\bn+\bid) \, \ba$ and its adjoint $f(\bn) \, \ba\daag = \ba\daag\, f(\bn+\bid)$ for any function $f: \mathbb{N} \mapsto \mathbb{C}$. Equation \eqref{eq:Hc} essentially expresses that atomic state can raise from $\q{g}$ to $\q{e}$ or from $\q{e}$ to $\q{m}$ by absorbing one photon from the field, or fall inversely by releasing one photon. The propagator ${\bf U}$, expressing the transformation that the joint atom-field state undergoes during interaction, follows the Schr\"odinger equation
\begin{equation}\label{eq:Uevaux}
	\tfrac{d}{dt} {\bf U}(t) = -i \, (\, H_f + H_a + H_c \,)\, {\bf U}(t)
\end{equation}
with initial condition ${\bf U}(t_0)=\bid$ the identity operator. A standard change of variables $\q{\psi}\, \rightarrow \, e^{i H_f t} \q{\psi}$ on field state and $(\q{g},\,\q{e},\,\q{m}) \; \rightarrow \; (e^{-i\omega_f}\q{g},\,\q{e},\,e^{i\omega_f}\q{m})$ on atomic states leads to `interaction coordinates'. In these coordinates the propagator follows
\begin{equation}\label{eq:Uev}
	\tfrac{d}{dt} {\bf U}(t) = -i \, H_{JC} \, {\bf U}(t)
\end{equation}
where the \emph{Jaynes-Cummings} Hamiltonian writes
\begin{equation}\label{eq:HJC}
	H_{JC}(u) = (\overline{\Delta}_m - u) \,\q{m}\qd{m} - (\overline{\Delta}_g + u) \,\q{g}\qd{g} + H_c
\end{equation}
with $\overline{\Delta}_m = \overline{\omega}_m - \omega_f$ and $\overline{\Delta}_g = \overline{\omega}_g - \omega_f$. The goal being to allow separate interactions of the two atomic transitions with the field, we assume that the $\q{m}$ and $\q{g}$ levels are attributed sufficiently different frequencies in $H_{JC}$, i.e.~$\vert \overline{\Delta}_g+\overline{\Delta}_m \vert =: \overline{\Delta} \gg \Omega$. Note that $H_{JC}$ acts in parallel on a set of decoupled subspaces spanned by $(\q{g}\otimes \q{n+1},\,\q{e}\otimes\q{n},\,\q{m}\otimes \q{n-1})$. The associated matrix operators are thus block-diagonal with blocks of size 3$\times$3 at most, which facilitates analysis.

The atoms are sent one after the other, every $T_s$ seconds, to undergo the same interaction with the field. We denote the field mode density operator just before interacting with the $(k+1)$th atom by $\rho_k$. The initial state $\vert u_\text{at} \rangle \in \mathbb{C}^3$ of the atoms can be chosen, as well as the Stark detuning signal $u(t)$ during interaction time $[0,T]$, with $T\leq T_s$. Denote $\bU_T$ the solution at time $T$ of \eqref{eq:Uev} with $\bU(0)=\bid$ and with the chosen $u(t)$, governing a time-varying $H_{JC}$ in \eqref{eq:HJC}. Then the atom-field joint state just after the $k+1$th interaction is given by $\bU_T \,(\rho_k \otimes \q{u_\text{at}}\qd{u_\text{at}})\, \bU_T$. This generally corresponds to an entangled situation. Since we do not measure the final atomic state, the \emph{expected} field evolution follows the Kraus map
\begin{equation}\label{eq:Kraus}
  \rho_{k+1} = \Phi(\rho_k) = M_g \rho_k M_g\daag + M_e \rho_k M_e\daag + M_m \rho_k M_m\daag	\, ,
\end{equation}
where the operators $M_g$, $M_e$, $M_m$ acting only on the field mode are identified from $\bU_T$ and $\q{u_\text{at}}$:  $\forall \q\psi\in\mathcal{H}$,
$$
\bU_T ~\q{\psi}\q{u_\text{at}}= M_g\q{\psi} ~\q{g} +M_e\q{\psi} ~\q{e}+M_m\q{\psi} ~\q{m}
.
$$
 The goal of open-loop stabilization by reservoir engineering is to select $u(t)$ and $\q{u_\text{at}}$ such that the dynamics \eqref{eq:Kraus} asymptotically stabilize a target pure state $\rho_k \rightarrow \rho_\infty = \q{\psi_\infty}\qd{\psi_\infty}$.


\section{Control design}\label{sec:control}

Our objective is to stabilize a given Fock state $\q{\psi_\infty} = \q{\targ}$.
\cite{WaltherTS} have noted that in absence of the $\q{m}$ level (i.e.~$\vert \overline{\Delta}_m - u \vert \gg \Omega$), a single atomic transition $(\q{g},\q{e})$ in perfect resonant interaction with the field (i.e.~$\overline{\Delta}_g + u =0$) allows to ``trap'' states below $\q{\targ}$. The corresponding propagator (obtained by direct integration of \eqref{eq:Uev} which is then 2$\times$2 block-diagonal) writes
\begin{eqnarray}
\label{eq:Uresonant} 	
 \bU_r & = & \cos (\tfrac{\theta_r \sqrt{\bn}}{2})\,\q{g}\qd{g}  +  \cos (\tfrac{\theta_r \sqrt{\bn+\bid}}{2})\, \q{e}\qd{e} \\
\nonumber  & & -  \ba \frac{\sin (\tfrac{\theta_r \sqrt{\bn}}{2})}{\sqrt{\bn}} \, \q{e}\qd{g} \, +   \frac{\sin (\tfrac{\theta_r \sqrt{\bn}}{2})}{\sqrt{\bn}} \, \ba\daag\, \q{g}\qd{e} \;,
\end{eqnarray}
where $\theta_r = t_r \Omega$ is the interaction strength integrated over chosen interaction time $t_r$. Then taking $\theta_r = 2\pi / \sqrt{\targ+1}$, the operator in \eqref{eq:Uresonant} implies no exchange between joint state components $\q{g}\otimes\q{\targ+1}$ and $\q{e}\otimes\q{\targ}$. The subspace $\mathcal{H}_0^{\targ}$ spanned by Fock states $\q{0},\q{1},...\q{\targ}$ then remains decoupled from the rest of the field Hilbert space throughout reservoir action. Now take $ \vert u_\text{at} \rangle=\q{e}$, thus $M_g=\frac{\sin (\tfrac{\theta_r \sqrt{\bn}}{2})}{\sqrt{\bn}} \, \ba\daag$, $M_e=\cos (\tfrac{\theta_r \sqrt{\bn+\bid}}{2})$ and $M_m=0$. Then $\rho_k$ following~\eqref{eq:Kraus} starting from $\rho_0$ with support in $\mathcal{H}_0^{\targ}$ converges to the trapping state $\q{\targ}\qd{\targ}$ \cite[page~210]{HRBook}. But if the support of $\rho_0$ is in $\mathcal{H}_{\targ+1}^{4\targ+3}$, then $\rho_k$ converges to $\q{4\targ+3}\qd{4\targ+3}$ as the field gets continuously excited. Therefore any fraction of density that is pushed above $\q{\targ}$ is lost away to high photon numbers. Since perturbations will always induce transitions between nearby energy states, this makes the method of \cite{WaltherTS} unusable in practice as it leaves equilibrium $\q{\targ}$ unstable in important directions. A robust stabilization method should enlarge the basin of attraction of $\q{\targ}\qd{\targ}$ to include at least any $\rho_0$ with support in $\mathcal{H}_{n_1}^{n_2}$ for some  $n_1 < \targ < n_2$.

We achieve such stabilization with $n_1=0$ and $n_2=4\targ+3$ (see Theorem~\ref{thm:conv}) by exploiting the possibility of varying $u(.)$ during the interaction. Specifically, we take
\begin{equation}\label{eq:usig}
	u(t) = \left\{
	\begin{array}{ll}
	     -\overline{\Delta}_g &  \text{ for } t \in [0,(T-t_s)/2]\\
		 \overline{\Delta}_m & \text{ for } t \in [(T-t_s)/2,(T+t_s)/2]\\
		 -\overline{\Delta}_g & \text{ for } t \in [(T+t_s)/2,T] \, . \\		
	\end{array}
	\right.
\end{equation}
Fast set-point changes can indeed suitably be experimentally implemented. The switching time $t_s$ and overall interaction time $T$ will be tuned to optimize operation. We still take $\q{u_\text{at}} = \q{e}$ as initial atomic state.

The propagator $\bU_T$, solution of \eqref{eq:Uevaux} with $u$ given by \eqref{eq:usig}, readily writes
$\bU_T = \bU_1 \, \bU_2 \, \bU_1 $
where $\bU_1 = \text{exp}[-i H_{JC}(-\overline{\Delta}_g)\, (T-t_s)/2]$ is the solution at $t=(T-t_s)/2$ of \eqref{eq:Uevaux} starting at $t_0=0$, with constant $u = -\overline{\Delta}_g$; and
$\bU_2 = \text{exp}[-i H_{JC}(\overline{\Delta}_m)\, t_s]$ is the solution at $t=t_s$ of \eqref{eq:Uevaux} starting at $t_0=0$, with constant $u = \overline{\Delta}_m$. We compute those operators using quantum Hamiltonian perturbation theory on the decoupled subspaces spanned by $(\q{g}\otimes \q{n+1},\,\q{e}\otimes\q{n},\,\q{m}\otimes\q{n-1})$ and neglecting terms of order $\Omega/\overline{\Delta} \ll 1$. Up to this approximation, the states $\q{m}\otimes\q{n-1}$ (resp.~$\q{g}\otimes \q{n+1}$) remain decoupled from the rest for $\bU_1$ (resp.~$\bU_2$). One gets:
\begin{multline}
\label{eq:UTnow}  \bU_T \q{e} = \overbrace{\ba^\dagger\, \left( \, e^{i\overline{\Delta}t_s} + \cos\tfrac{\theta_2\sqrt{\bn}}{2} \, \right) \;\frac{\sin(\theta_1 \sqrt{\bn+\bid})}{2\sqrt{\bn+\bid}}}^{M_g} \; \q{g} \phantom{k r phto tneoa tno \omega}\\
 + \overbrace{\left( \, \cos^2\tfrac{\theta_1 \sqrt{\bn+\bid}}{2} \, \cos\tfrac{\theta_2\sqrt{\bn}}{2} - e^{i\overline{\Delta}t_s} \, \sin^2\tfrac{\theta_1 \sqrt{\bn+\bid}}{2} \, \right)}^{M_e}\; \q{e}\\
 -\overbrace{ e^{-i\overline{\Delta}(T\minou t_s)/2} \; \ba \, \frac{\sin\tfrac{\theta_2\sqrt{\bn}}{2}}{\sqrt{\bn}} \, \cos\tfrac{\theta_1\sqrt{\bn+\bid}}{2}}^{M_m} \; \q{m}
\end{multline}
with $\theta_1 = \Omega\, (T-t_s)/2$ and $\theta_2 = \Omega\, t_s$. To make the target field state $\q{\targ}$ invariant under the Kraus map \eqref{eq:Kraus}, we make it invariant under $\bU_T \q{e}$. This is achieved with the trapping-like condition:
\begin{equation}\label{eq:conds}
 (T-t_s) = \frac{2\pi}{\Omega\sqrt{\targ+1}} \quad \Leftrightarrow \quad \theta_1 = \frac{\pi}{\sqrt{\targ+1}} \; .
\end{equation}
For simplicity, we take $\overline{\Delta}\,t_s = 0$ nominally\footnote{To this end, the value of $\overline{\Delta}$ can be slightly tuned by adjusting the trap that governs field mode frequency $\omega_f$. Indeed since $\overline{\Delta} \gg \Omega$, a small shift in $\overline{\Delta}$ allows to sensibly tune $\overline{\Delta}\,t_s$ for all $t_s$ that yield non-negligible values of $\theta_2 = \Omega t_s$.}. The value of $\overline{\Delta}(T-t_s)$ is irrelevant as it drops out of the Kraus map for the field evolution. The value of $\theta_2$ remains to be fixed.


\section{Convergence analysis} \label{sec:conv}

Condition \eqref{eq:conds} makes the subspace $\mathcal{H}_{0}^{(4\targ+3)}$ invariant by $M_g$, $M_e$ and $M_m$. We denote $\mathcal{P}_{0}^{(4\targ+3)}$ the orthogonal projection onto $\mathcal{H}_{0}^{(4\targ+3)}$. Consider a candidate Lyapunov function of the form $V(\rho)= \text{trace}(f_\bn \, \rho)$, for some function $f(n)$ to be determined. Then \eqref{eq:Kraus} and \eqref{eq:UTnow} lead to
\begin{multline*}
V(\Phi(\rho)) - V(\rho) =\\
\text{trace}\left(\, \rho \; \cos^2(\tfrac{\alpha_\bn}{2}) \sin^2(\beta_\bn) \, \left(\, f(\bn-\bid) - f(\bn) \,\right) \, \vphantom{\, \rho \; \sin^2(\alpha_\bn) \cos^4(\tfrac{\beta_\bn}{2}) \, \left(\, f(\bn+\bid) - f(\bn) \,\right) \,}\right)\\
+ \text{trace}\left(\, \rho \; \sin^2(\alpha_\bn) \cos^4(\tfrac{\beta_\bn}{2}) \, \left(\, f(\bn+\bid) - f(\bn) \,\right) \, \right)
\end{multline*}
with $\alpha_\bn =  \pi \sqrt{\tfrac{\bn+\bid}{\targ+1}}$ and $\beta_\bn = \theta_2 \sqrt{\bn}/2$. Note that the last line of the above equation vanishes for $\rho = \q{4\targ+3}\qd{4\targ+3}$ and any function $f_\bn$, reflecting the decoupling of $\mathcal{H}_{0}^{(4\targ+3)}$ from the remainder of the Hilbert space. To formally restrict ourselves to $\mathcal{H}_{0}^{(4\targ+3)}$, we can take $f(n) = f(4\targ+3)$ for all $n>4\targ+3$, such that $f(\bn-\bid)-f(\bn)$ and $f(\bn+\bid)-f(\bn)$ vanish on $\mathcal{H}_{(4\targ+4)}^{+\infty}$. Further take $\eta\in(0,1)$,  $f(\targ) = 0$, $f(\targ+1) = f(\targ-1) = 1$ and set
\begin{eqnarray}
\nonumber f(n\!-\!1) & = & f(n) + \eta \, \sin^2 \tfrac{\alpha_n}{2} \cos^2 \tfrac{\beta_n}{2}\, (f(n)-f(n\!+\!1)) \\
\nonumber & & \text{for } \; 0<n<\targ \; , \\
\label{eq:fdef1} f(n\!+\!1) & = & f(n) + \eta \, \sin^2 \tfrac{\beta_n}{2}\, (f(n)-f(n\!-\!1)) \\
\nonumber & & \text{for } \; \targ<n<4\targ+3 \; .
\end{eqnarray}
Then  $ V(\Phi(\rho)) - V(\rho) = \text{trace}(q_\bn\, \rho) $ with $q_n=0$ for $n=\targ$,  $n>4\targ+3$, and
\begin{eqnarray}
\nonumber
q_n & = & \sin^2 \alpha_n  \cos^4 \tfrac{\beta_n}{2}\, (\eta\sin^2 \tfrac{\beta_n}{2} - 1) \, (f(n)\minou f(n\!+\!1)) \\
\label{eq:bn} & & \text{for } \; 0\leq n<\targ \; , \\
\nonumber q_n & = & \sin^2 \beta_n \cos^2 \tfrac{\alpha_n}{2} \, (\eta\sin^2\tfrac{\alpha_n}{2}\cos^2\tfrac{\beta_n}{2}- 1) \, (f(n)\minou f(n\minou 1)) \\
\nonumber & & \text{for } \; \targ < n \leq 4\targ+3 \; .
\end{eqnarray}
We then have the following convergence result.\\

\begin{thm} \label{thm:conv} Consider the dynamics~\eqref{eq:Kraus} where $M_g$, $M_e$ and $M_m$ are defined by~\eqref{eq:UTnow} with ~\eqref{eq:conds}, and its restriction to  density operators $\rho$ with support in~$\mathcal{H}_{0}^{(4\targ+3)}$. Assume that $\theta_2 \neq  k \pi / \sqrt{n}$ for all $(n,k)\in\{1,...,4\targ+3\}\times\mathbb{N}$. Then $V(\rho)$ built with \eqref{eq:fdef1} is a strict Lyapunov function: for any $\rho_0$ with support in~$\mathcal{H}_{0}^{(4\targ+3)}$, $\rho_k$ converges towards the fixed point $\rho_\infty = \q{\targ}\qd{\targ}$.
\end{thm}
\emph{Proof:} Thanks to the assumption on $\theta_2$, \eqref{eq:fdef1} implies $f(n) > f(n-1)$ for $\targ < n \leq 4\targ+3$ and $f(n) > f(n+1)$ for $0 < n < \targ$. Then the same assumption ensures that $q_n$ is \emph{strictly} negative for all $n \leq 4\targ+3$ except $n=\targ$. Writing $\rho = \sum_j \, p_j \q{\psi_j} \qd{\psi_j}$ (spectral decomposition with $p_j>0$, $\sum_j p_j=1$)  then yields
$$V(\Phi(\rho))-V(\rho) = \sum_j \, p_j \qd{\psi_j} \, q_\bn \, \q{\psi_j} < 0$$
unless $\mathcal{P}_{0}^{(4\targ+3)} \q{\psi_j} = s_j \, \q{\targ}$ with $s_j \in \mathbb{C}$ for all $j$, implying $\mathcal{P}_{0}^{(4\targ+3)} \rho \mathcal{P}_{0}^{(4\targ+3)} = p \q{\targ}\qd{\targ}$ for some $p \in \mathbb{R}$. Thus $V$ is a strict Lyapunov function in $\mathcal{H}_{0}^{(4\targ+3)}$ as stated. \hfill $\square$

\emph{Remark:} The above construction of $f(n)$ can in fact be extended to $n<9\targ+8$. For generic $\theta_2$, such that $\sin(2\beta_n)$ never vanishes, the resulting $V$ is then a non-strict Lyapunov function on the invariant Hilbert subspace $\mathcal{H}_{0}^{(9\targ+8)}$, with the Lasalle invariance principle ensuring convergence of $\rho$ to a $\rho_\infty$ supported on the subspace spanned by $\q{m},\q{9m+8}$. Considering even larger Hilbert spaces, one shows that $\rho$ converges to a state with nonzero population only on Fock states $\q{(2l+1)^2 (\targ+1) - 1}$ for $l$ integer. The experiments working in very low temperature environments ensure that decoherence usually pulls the field state towards vacuum (Fock state $\q{0}$) and depopulates high-number Fock states.


Tailored interaction~\eqref{eq:usig} with the atomic stream thus makes the otherwise isolated field converge to $\q{\targ}$ for virtually all $\theta_2 > 0$. This confirms the interest of this symmetric product operator approach. We next analyze how the latter can be optimized to strengthen convergence w.r.t.~external disturbances. We therefore add to the evolution model a typical relaxation disturbance, also called decoherence: evolution between consecutive atomic samples becomes
\begin{eqnarray}
	\label{eq:dissip} \rho_{k+1} & = & \Phi(\rho_k) - \tfrac{\Gamma^-}{2} [\bn \Phi(\rho_k) + \Phi(\rho_k) \bn - 2 \ba \Phi(\rho_k) \ba^\dagger]\\
	\nonumber & & - \tfrac{\Gamma^+}{2} [(\bn+\bid)\Phi(\rho_k) + \Phi(\rho_k)(\bn+\bid) - 2 \ba^\dagger \Phi(\rho_k) \ba ]
\end{eqnarray}
with $\Gamma^+ \!=\! \kappa n_{th} T_s \, \ll \, \Gamma^- \!=\! \kappa (1+n_{th}) T_s \, \ll \, 1$. The terms in $\Gamma^-$ and $\Gamma^+$ model interaction with a thermal environment that induces photon annihilation and creation respectively. Parameter $\kappa$ models coupling strength to this environment, while $n_{th}$ is the average number of thermal photons per mode in the environment. The invariant-subspace structure no longer rigorously holds, but we can invoke a physical argument to still truncate our computations at $n \leq 9\targ+8$, ensuring that the discretized model remains valid and all parameters in the following are well-behaved.

Writing the operator $\rho_k$ as a matrix with components $[\rho_k]_{a,b}$, for $a\geq 0$ and $b\geq 0$, the evolution \eqref{eq:dissip} couples component $[\rho_k]_{a,b}$ only to components $[\rho_k]_{a+l,b+l}$ i.e.~on the same diagonal of the matrix. To analyze how the fidelity $\qd{\targ} \rho_k \q{\targ}$ to target $\q{\targ}\qd{\targ}$ evolves, we may thus reduce our investigation to the vector $r_k$ containing the principal diagonal of $\rho_k$, that is $[r_k]_{a} = [\rho_k]_{a,a}$ for $a=0,1,2,...\, $. Evolution \eqref{eq:dissip} yields
$$r_{k+1} = B \cdot A \cdot r_k$$
with $A$ and $B$ tridiagonal real positive non-symmetric matrices, representing reservoir and decoherence respectively. The upper, lower and principal diagonals of $B$ (respectively $A$) have elements $b_n = \Gamma^- \, n$ (resp.~$d_n = \sin^2 \beta_n \cos^2 \tfrac{\alpha_n}{2}$) for $n\in\{1,2,...\}$; $c_n = \Gamma^+ \, n$ (resp.~$e_n=\sin^2\alpha_n \cos^4 \tfrac{\beta_n}{2}$) for $n\in\{0,1,2,...\}$; and $1-c_n-b_n$ (resp.~$1-d_n-e_n$). $A$ and $B$ are thus column-stochastic, reflecting conservation of $\text{trace}(\rho) = \mathbf{1}^T r$ where $\mathbf{1}$ is the vector of all ones. Therefore $B \cdot A$ has at least one eigenvalue 1. The corresponding $1$-eigenvector $r_*$ characterizes photon populations for a field pointer state under reservoir and decoherence.

We try to approximate it by viewing $B\cdot A $ as a small perturbation of $A$. Denote $x^0$ the $r$-vector corresponding to $\q{\targ}\qd	{\targ}$, let $R^0 = A-\bid$ and $R^1 = (B-\bid)(A-\bid)+(B-\bid)$. Then we know that $R^0 x^0 = 0$ with $\mathbf{1}^T x^0 = 1$, i.e.~$x^0$ is the steady-state solution $r_*$ when $B=\bid$ (no perturbation). In presence of a thermal environment, we get $r_*=x^0+x^1$ solution of
\begin{equation}\label{eq:tosolve}
	(R^0+R^1)(x^0+x^1) = 0, \quad \mathbf{1}^T x^1 = 0 \, .
\end{equation}\\

\begin{prop} \label{prop:decoherence}
Approximating $R^1\, x^1 \approx 0$, problem \eqref{eq:tosolve} can be explicitly solved. It gives $1+x^1_{\targ}$ as estimated fidelity of $\rho_\infty$ to our goal, where
$$-x^1_{\targ} = \frac{b_{\targ}}{e_{\targ-1}}\, \sum_{m=1}^{\targ} \, \prod_{l=2}^m \, \frac{d_{\targ-l+1}}{e_{\targ-l}} \, + \frac{c_{\targ}}{d_{\targ+1}}\, \sum_{m>0} \,
\prod_{l=2}^m \, \frac{e_{\targ+l-1}}{d_{\targ+l}} \, .$$
An approximation error of order $(x^1_{\targ})^2$ is expected.
\end{prop}
\emph{Proof:} Solving \eqref{eq:tosolve} approximated  (note that $d_{\targ}=e_{\targ}=0$) fixes the components of $x^1$, except $x^1_{\targ}$ which remains free:
\begin{eqnarray*}
	x^1_{\targ-1} = b_{\targ} / e_{\targ-1} & \;\; ; \;\;\;\; & x^1_n = \frac{d_{n+1}}{e_n} x^1_{n+1} \; \forall\, n<\targ-1\; ; \\
	x^1_{\targ+1} = c_{\targ} / d_{\targ+1} & \;\; ; \;\;\;\; & x^1_n = \frac{e_{n-1}}{d_n} x^1_{n-1} \; \forall\, n>\targ+1\; .
\end{eqnarray*}
Then \eqref{eq:tosolve} determines $x^1_{\targ}<0$. \hfill $\square$

Proposition 2 can be used to optimize $\theta_2$ for maximal fidelity in presence of decoherence, see next Section. Note that $\theta_2=0$ (the case of \cite{WaltherTS}) would yield $d_n = 0$ for all $n$. The small-$x_1$ approximation would then lead to an invalid recurrence, suggesting that decoherence leads to large $x_1$ i.e.~the method of \cite{WaltherTS} is poorly robust.


\section{Simulations} \label{sec:sim}

\begin{figure}\begin{center}
	\setlength{\unitlength}{1mm}
	\begin{picture}(80,45)
	\put(5,0){\includegraphics[height=47mm]{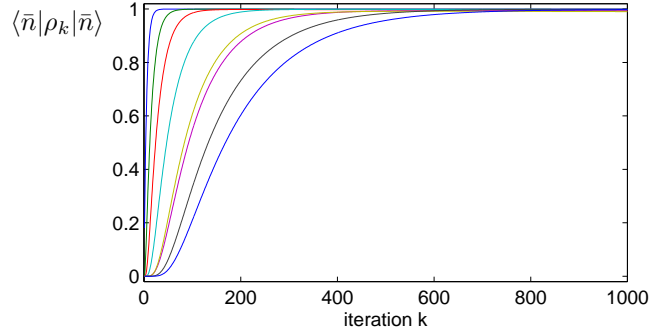}}
	\put(-2,40){$\qd{\targ} \rho_k \q{\targ}$}
    \end{picture}
	\caption{Fidelity $\qd{\targ} \rho_k \q{\targ}$ of $\rho_k$ to target state $\q{\targ}$ as a function of atomic interaction $k$, for $\targ=1,2,...,8$ (going from top to bottom curve) and $\Gamma^- = \Gamma^+ = 0$ i.e.~no environmental disturbance.}\label{fig1}
\end{center}\end{figure}

For the simulations we consider realistic parameters corresponding to the cavity quantum electrodynamics setup at Ecole Normale Sup\'erieure, Paris. See e.g.~\cite{HRBook,Deleglise08} for detailed explanations. Vacuum Rabi frequency is given by $\Omega/2\pi=50$~kHz and we take $\overline{\Delta} \approx 100 \Omega$. We numerically compute propagators for the exact interaction \eqref{eq:HJC} and $u$ given by \eqref{eq:usig}. For each simulation we slightly adapt $\overline{\Delta}$ to optimize fidelity, see footnote 2. Evolution of $\rho_k$ is computed starting from a vacuum initial state $\rho_0 = \q{0}\qd{0}$. Figure~\ref{fig1} illustrates the good working principle of the method, despite our approximate reasoning: $\rho_{k+1} = \Phi(\rho_k)$ converges essentially exactly to $\rho_\infty = \q{\targ}\qd{\targ}$ for all $\targ \in \{1,2,...,8\}$. The simulations are run with an arbitrary $\theta_2 = 1/\sqrt{\targ}$.

\begin{figure}\begin{center}
	\includegraphics[height=61mm,trim=2cm 0cm 0cm 1cm,clip=true]{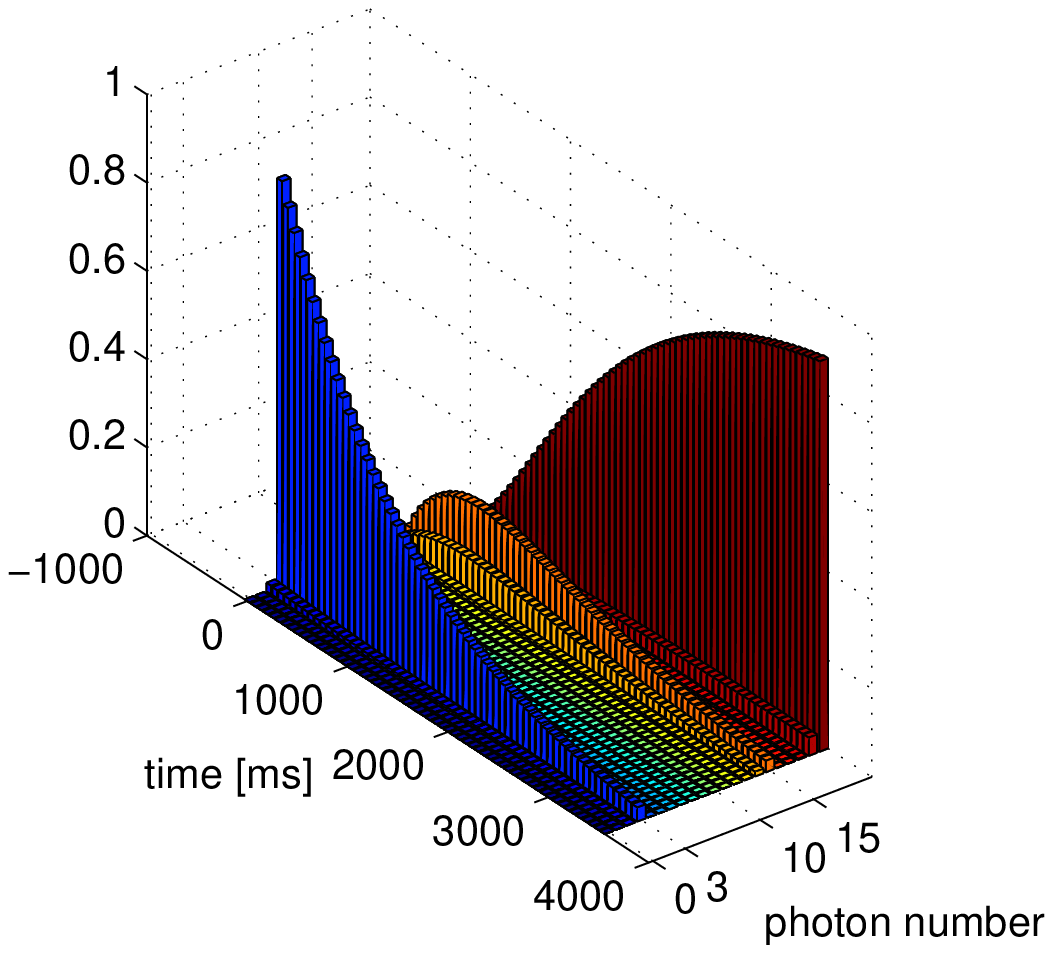}
	\includegraphics[height=61mm,trim=2cm 0cm 0cm 1cm,clip=true]{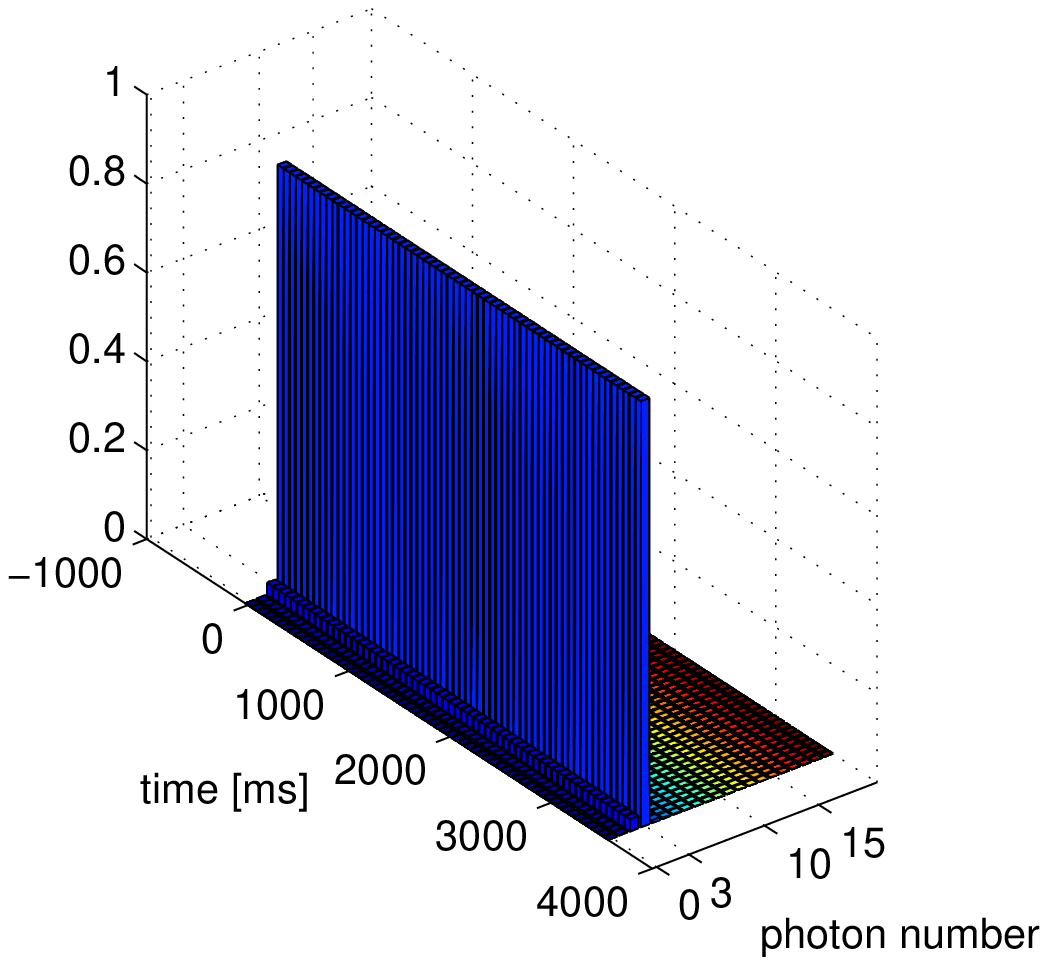}
		\caption{Evolution of the diagonal elements of $\rho_k$ in presence of a disturbing environment like in \eqref{eq:dissip}, with the reservoir of \cite{WaltherTS} (top) and with our symmetric-interaction reservoir (bottom), for $\targ=3$. Graduation is given as a function of time =  $k\; \cdot$ 60~$\mu$s.}\label{fig2}
\end{center}\end{figure}

\begin{figure}\begin{center}
	\setlength{\unitlength}{1mm}
	\begin{picture}(80,51)
	\put(-2,0){\includegraphics[height=54mm]{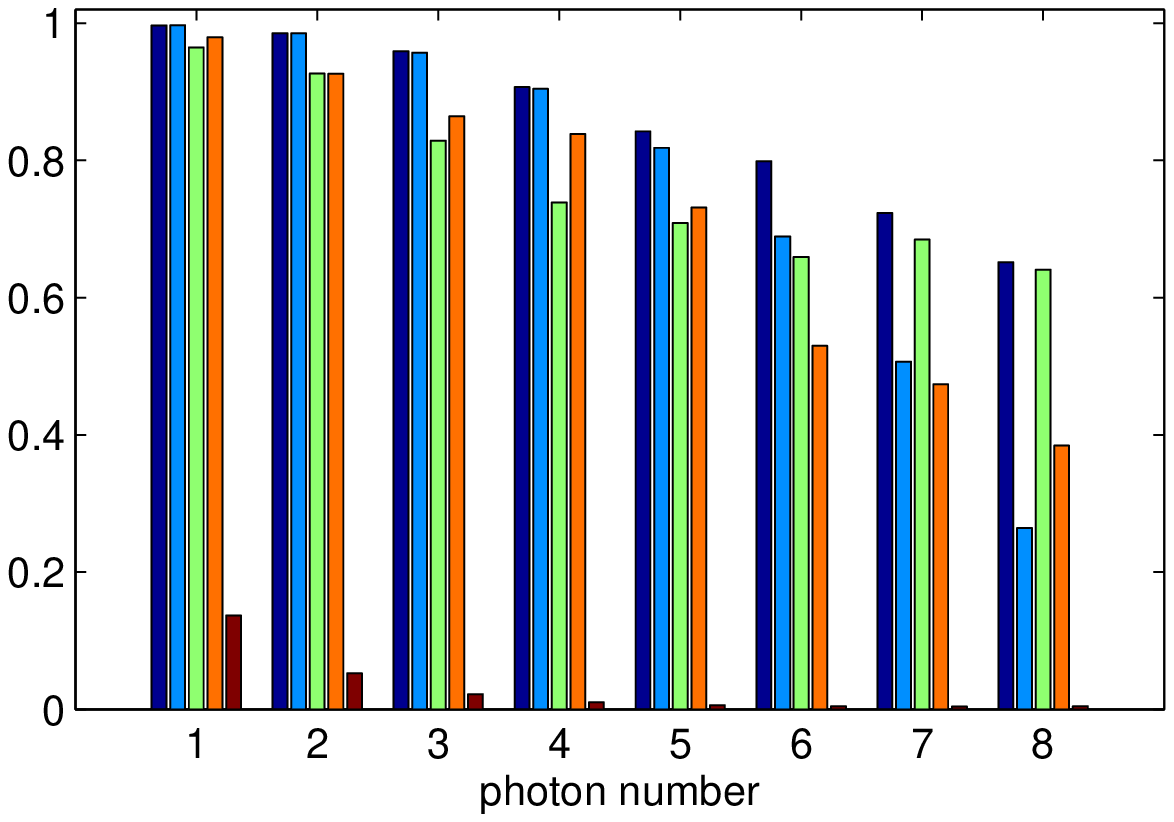}}
	\put(61,45){$\qd{\targ} \rho_k \q{\targ}$}
    \end{picture}
	\caption{Fidelity $\qd{\targ} \rho_\infty \q{\targ}$ of the steady state $\rho_\infty$ in presence of a disturbing environment to target states $\q{\targ}$, for $\targ=1,2,...,8$. The left-most bars represent fidelity with our simulated reservoir, the second are the values predicted by Proposition 2, and the right-most ones are obtained after $4$~s with the reservoir of \cite{WaltherTS}. The third and fourth bars from left correspond to the same conditions as the first ones, but incorporating a systematic error of $-2\%$ and $+2\%$ respectively on $\theta_1$.}\label{fig3}
\end{center}\end{figure}

We next add decoherence. We take $1/\kappa = 0.1\,s$ and cryogenic $n_{th} = 0.05$, corresponding to current high-standard experiments. This environment-induced decoherence is in competition with our reservoir strength, that is mainly the time between consecutive atoms. The latter can realistically\footnote{Although the required values of $\theta_1$ and $\theta_2$ allow smaller $T$, we are experimentally limited to periods of the order of $T_s$.} be set to $T_s=60$~$\mu$s, but there is only a $0.3$ probability that an interacting atom is indeed present when trying to send one. The expected evolution from $\rho_k$ to $\rho_{k+1}$ is thus similar to (a less approximated version of) \eqref{eq:dissip} except that operator $0.7\, \bid + 0.3\, \Phi$ replaces $\Phi$. Figure~\ref{fig2} represents the evolution of the diagonal elements of $\rho$ with the reservoir of \cite{WaltherTS} (top) and with our symmetric-interaction reservoir (bottom), for a target $\targ=3$. For the latter, after the $\q{3}$ state has built up, it progressively drives away towards $\q{15}	 = \q{4\targ+3}$ and further to higher photon numbers (here accumulating where our Hilbert space is truncated). In contrast our scheme clearly stabilizes a state close to $\q{3}$. Achievable steady-state fidelities are represented on Fig.~\ref{fig3}. We also show the values estimated from Proposition 2. They agree with simulations within the expected approximation error, confirming our theoretical analysis. The selected optimal values of $\theta_2$ seem to satisfy $\theta_2 \sqrt{\targ} \approx 3\pi/4$.

Finally, we consider errors in model parameters as another important robustness criterion. Small variations in $\theta_2$ appear to give no detectable effect, indicating a rather flat optimum in $\theta_2$. Absolute errors of $\pm \pi/8$ on $\overline{\Delta} t_s$ also barely affect fidelity (less than $1\%$). The fidelities obtained with our reservoir subject to errors in $\theta_1$  (and still in presence of decoherence) are represented for each $\targ$ by the 3rd and 4th bars from left on Fig.~\ref{fig3}. Relative errors of $\pm 2\%$ have a detectable but tolerable effect; especially, setting $\theta_1$ in an interval centered slightly below the ideal value seems to be a good robustness compromise. The same $\pm 2\%$ error on $\theta_1$ would be drastically detrimental with the scheme of \cite{WaltherTS}: even in absence of a disturbing environment ($\Gamma^- = \Gamma^+ = 0$) the fidelity to $\q{\targ}$ would then quickly decrease towards zero, dropping to about $0.15$ already after $0.1$~s and below $0.01$ after $0.25$~s. Overall our construction of atom-field interaction as a product of non-commuting transformations thus leads to a significantly more robust stabilization scheme.

\section{Conclusion}

This paper proposes an `engineered reservoir' to stabilize Fock states of a quantum harmonic oscillator thanks to its interaction with a stream of three-level systems.  We use a single control signal to tailor a time-varying Hamiltonian interaction. We prove that the resulting propagator yields a Kraus map that robustly `traps' the spring state on a desired Fock state. This seems to be a potentially practical open-loop alternative to measurement-based feedback for achieving high-fidelity stabilization of Fock states. The proposed method admits several variations for experimental implementation. Among others, a similar stabilizing effect is obtained by using a stream of two-level atoms each undergoing one of two different interactions. Given the practical possibilities shown in the present and previous papers, future work could further investigate systematic methods for designing products of interactions that robustly stabilize quantum states.


\begin{ack}
The authors thank Michel Brune, Igor Dotsenko and Jean-Michel Raimond from Ecole Normale Sup\'{e}rieure for useful discussions.
\end{ack}

\bibliographystyle{alpha}
\bibliography{QifacBib}

\end{document}